\documentclass[11pt]{article}
\usepackage{moriond,epsfig}

\def\be{\begin{equation}}
\def\ee{\end{equation}}
\def\bea{\begin{eqnarray}}
\def\eea{\end{eqnarray}}

\begin{document}
\vspace*{4cm}
\title{SEMILEPTONIC AND LEPTONIC CHARM DECAYS AT CLEO-C}

\author{ WERNER M. SUN (for the CLEO Collaboration) }

\address{Cornell University, Ithaca, New York 14853, USA}

\maketitle\abstracts{
Using $e^+e^-$ collision data in the $\sqrt{s}\approx 4$ GeV energy region,
CLEO-c has made extensive studies of semileptonic and leptonic decays of
the $D^0$, $D^+$, and $D_s^+$ charmed mesons.  We report recent measurements of
absolute branching fractions, form factors, and decay constants that serve
as precision tests of theoretical calculations.
}

\section{Introduction}\label{sec:intro}

In the past decade, the arrival of CLEO-c and now BES-III has brought a wealth
of new experimental data on charm physics.  These pristine data samples collected
at charm threshold complement the high statistics charm
samples collected at $B$-factories and hadron colliders.
This experimental renaissance is matched by the maturity of Lattice QCD (LQCD).
Testing LQCD calculations of charm form
factors and decay constants against measurements by CLEO-c and BES-III
validates its use in other related systems, such as $B$ decays.

The CLEO-c data samples discussed in this article were produced
at the Cornell Electron Storage Ring, with $e^+e^-$ collisions occuring at
two center-of-mass energies in the charm threshold region.  Here,
charm mesons are pair-produced nearly at rest in the lab frame, and the particle
multiplicities are ${\cal O}(10)$ per event.
The first data sample consists of 818 ${\rm pb}^{-1}$ produced on the
$\psi(3770)$ resonance, corresponding to $3.0\times 10^6$ $D^0\bar D^0$
events and $2.4\times 10^6$ $D^+D^-$ events.
The second sample consists of 600 ${\rm pb}^{-1}$ taken near
$\sqrt{s}=4170$ MeV, corresponding to $5.5\times 10^5$ $D_s^{*\pm}D_s^\mp$
events.  In the remainder of this article, sums over charge
conjugate states are implied.

To reduce backgrounds, we tag one of the two $D$ mesons in each event via
full reconstruction.  In this way, we infer not only the presence of a
second (signal) $D$ meson, but also its flavor and charge.  At the $\psi(3770)$,
$D$ and $\bar D$ are produced with no extra particles, while
at $\sqrt{s}=4170$ MeV, pairs of $D_s$ mesons are produced with a transition
$\gamma$ or $\pi^0$ from the $D_s^*$ decay.  We fully reconstruct
10--15\% of all $D^0$/$D^+$ decays and approximately 6\% of all $D_s$ decays.

We select electron tracks based on a multivariate discriminant that makes use
of energy deposited in the electromagnetic calorimeter (compared to the
track momentum),
ionization energy loss of the track in the drift chamber ($dE/dx$),
and information from the Ring Imaging Cherenkov counter (RICH).
Muons are not explicitly identified; instead, we veto tracks that are
associated with calorimeter deposits ({\it i.e.}, inconsistent with
minimum-ionizing muons).  We also veto
charged kaons identified by $dE/dx$ and the RICH.

One signature of semileptonic decays ($D_{(s)}\to X\ell\nu$) and
leptonic decays ($D_{(s)}^+\to\ell^+\nu$) is the presence
of a weakly-interacting neutrino.
We identify events containing a single neutrino by exploiting the hermeticity
of the CLEO-c detector.  We combine our knowledge of
the $e^+e^-$ beam parameters with
a $D$ tag and the visible candidates from the
(semi)leptonic signal $D$ decay to form the missing four-momentum of the event.
For signal events, the invariant mass of this four-momentum is consistent with
the neutrino mass of (approximately) zero.

Absolute branching fractions are obtained by dividing signal event
yields by tag yields after efficiency corrections.  For leptonic decays,
these branching fractions lead to a determination of the $D^+$ and $D_s^+$
decay constants $f_D$ and $f_{D_s}$ via
\begin{equation}
\Gamma(D_{(s)}\to\ell^+\nu_\ell) =
\frac{G_F^2 |V_{c\{d,s\}}|^2 f_{D_{(s)}}^2}{8\pi}
m_{D_{(s)}} m_\ell^2
\left( 1 - \frac{m_\ell^2}{m_D^2} \right)^2.
\end{equation}
For $D$ semileptonic decays, in addition to absolute branching fractions, we also
measure event yields in bins of $q^2$, the square of the virtual $W$
invariant mass.  The differential decay rates obtained from these yields
are related to the $D\to X$ form factors $f_+^X$ via
\begin{equation}
\frac{d\Gamma}{dq^2} =
\frac{G_F^2 |V_{c\{d,s\}}|^2 p_{K,\pi}^3}{24\pi^3}
| f_+^{X}(q^2) |^2.
\end{equation}

\section{Results}

\subsection{Semileptonic Decays}\label{subsec:prod}

CLEO-c results for $D^0\to\{K^-,\pi^-\}e^+\nu_e$ and
$D^+\to\{\bar K^0,\pi^0\}e^+\nu_e$ decays are based on two complementary analyses
using a 281 ${\rm pb}^{-1}$ subset of the $\psi(3770)$ dataset.
The first analysis~\cite{dslUntaggedPRL,dslUntaggedPRD} employs the tagging
technique described in
Section~\ref{sec:intro}.  The second analysis~\cite{dslTagged} does not require
a tag $D$
and instead infers the neutrino four-momentum from the {\it all} the visible
particles in an event.  This untagged analysis attains higher efficiency than
the tagged analysis, at the price of lower purity and larger systematic
uncertainties.  In averaging the results of these two analyses, we account for
sample overlap and correlated systematic uncertainties.

\begin{figure}[htb]
\begin{center}
\includegraphics*[height=5cm]{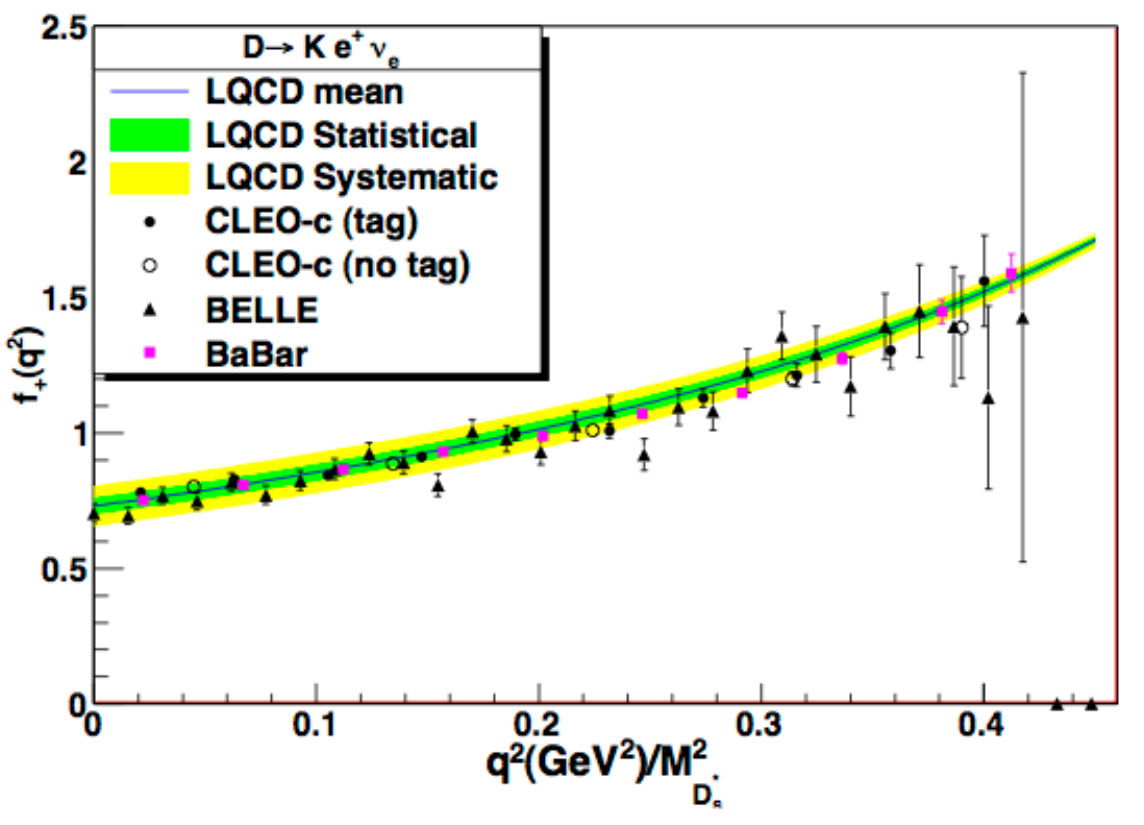}
\includegraphics*[height=5cm]{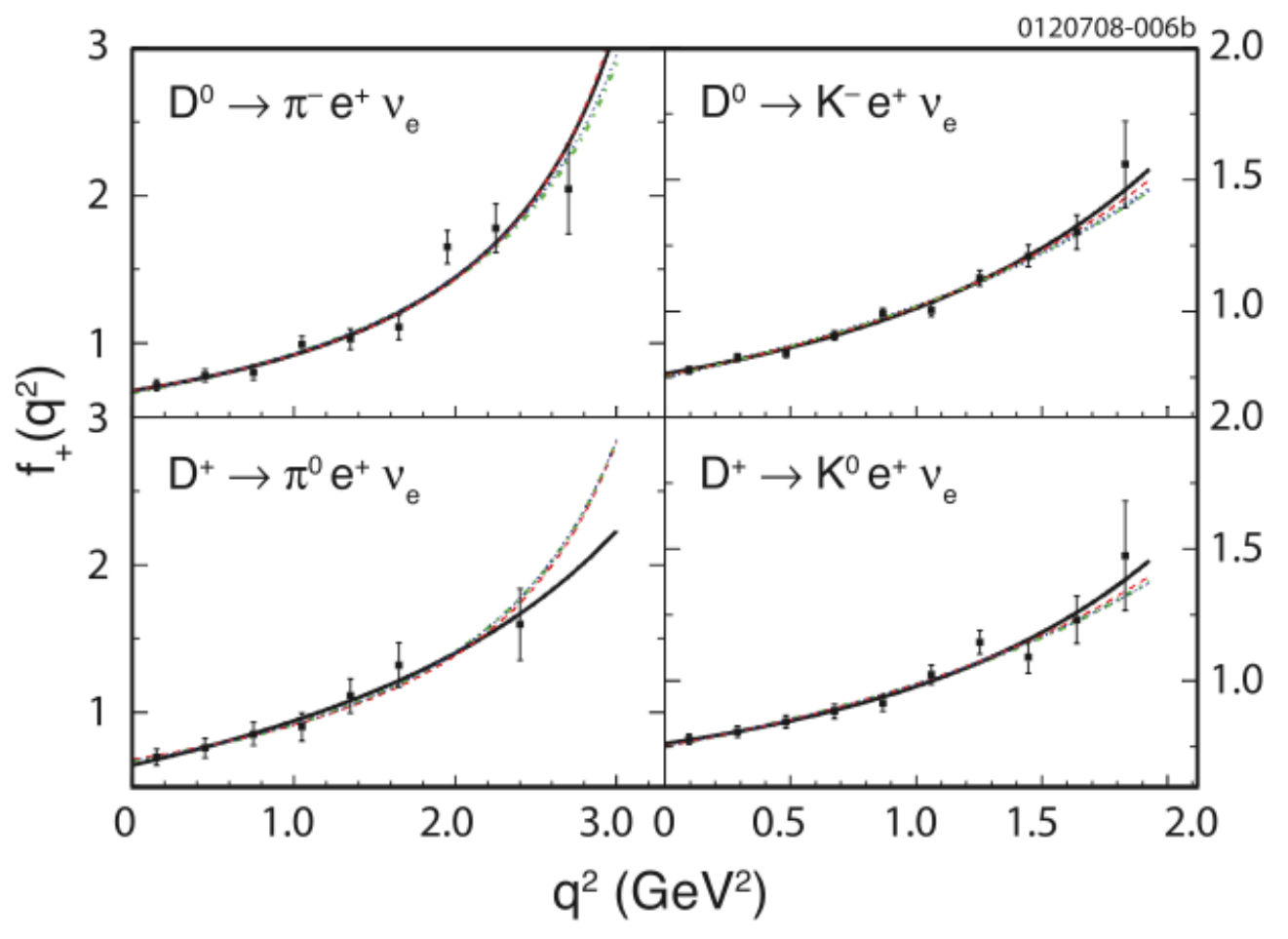}
\caption{
CLEO-c data for $f_+(q^2)$ compared to LQCD predictions 
(left) and fitted (right) to the simple (long dash) and modified (short dash) pole 
models and the two- (dot) and three-parameter (solid) series expansion.
}
\label{fig:ff}
\end{center}
\end{figure}

Branching fractions for the $D^0$ and $D^+$ semileptonic modes are reported
in Table~\ref{tab:brDSL}.  The precision of these measurements exceeds all
previous results.
In Figure~\ref{fig:ff}, we show the measured $d\Gamma/dq^2$ distributions
compared to LQCD~\cite{Aubin:2004ej} and
fitted to four models: two pole models~\cite{poleModel} of the form
$f_+(q^2) = f_+(0)/(1 - q^2/M_{\rm pole}^2)(1 - \alpha q^2/M_{\rm pole}^2)$,
with $\alpha = 0$ (simple) and $\alpha > 0$ (modified); and two- and
three-parameter forms of the series expansion discussed in
Refs.~\cite{series1,series2,series3,series4}.
All models are capable of describing the data,
although the pole model fits prefer unphysical pole masses~\cite{Hill:2006ub}.
By taking the LQCD value of $f_+^{K,\pi}(0)$~\cite{Follana:2007uv}, we also
obtain $|V_{cd}|=0.223\pm 0.008\pm 0.003\pm 0.023$ and
$|V_{cs}|=1.019\pm 0.010\pm 0.007\pm 0.106$, where the uncertainties are
statistical, experimental systematic, and from LQCD, respectively.

\begin{table}[t]
\caption{$D^{0/+}$ semileptonic branching fractions (\%).
Uncertainties are statistical and systematic, respectively.
\label{tab:brDSL}}
\vspace{0.4cm}
\begin{center}
\begin{tabular}{|l|c|c|c|}
\hline
Mode & Tagged & Untagged & Average \\
\hline
$D^0\to\pi^-e^+\nu_e$ & $0.308\pm 0.013\pm 0.004$ & $0.299\pm 0.011\pm 0.008$ &
	$0.304\pm 0.011\pm 0.005$ \\
$D^+\to\pi^0 e^+\nu_e$ & $0.379\pm 0.027\pm 0.023$ & $0.373\pm 0.022\pm 0.013$&
	$0.378\pm 0.020\pm 0.012$ \\
$D^0\to K^- e^+\nu_e$ & $3.60\pm 0.05\pm 0.05$ & $3.56\pm 0.03\pm 0.09$ &
	$3.60\pm 0.03\pm 0.06$ \\
$D^+\to \bar K^0 e^+\nu_e$ & $8.87\pm 0.17\pm 0.21$ & $8.53\pm 0.13\pm 0.23$ &
	$8.69\pm 0.12\pm 0.19$ \\
\hline
\end{tabular}
\end{center}
\end{table}

Results for $D_s^+$ semileptonic decays are based on
310 ${\rm pb}^{-1}$ at $\sqrt{s}=4170$ MeV using a tagging technique~\cite{dssl}.
Table~\ref{tab:brDsSL} shows first measurements of absolute $D_s^+$
semileptonic branching fractions and the first observations of the
Cabibbo-suppressed modes ($D_s^+\to K^{(*)0}e^+\nu_e$)
as well as $D_s^+\to f_0(980)e^+\nu_e$.

\begin{table}[t]
\caption{$D_s^+$ semileptonic branching fractions.
Uncertainties are statistical and systematic, respectively.
\label{tab:brDsSL}}
\vspace{0.4cm}
\begin{center}
\begin{tabular}{|l|c|}
\hline
Mode & ${\cal B}(\%)$ \\
\hline
$D_s^+\to\phi e^+\nu_e$ & $2.29\pm 0.37\pm 0.11$ \\
$D_s^+\to\eta e^+\nu_e$ & $2.48\pm 0.29\pm 0.13$ \\
$D_s^+\to\eta' e^+\nu_e$ & $0.91\pm 0.33\pm 0.05$ \\
$D_s^+\to K^0 e^+\nu_e$ & $0.37\pm 0.10\pm 0.02$ \\
$D_s^+\to K^{*0} e^+\nu_e$ & $0.18\pm 0.07\pm 0.01$ \\
$D_s^+\to f_0(\pi^+\pi^-) e^+\nu_e$ & $0.13\pm 0.04\pm 0.01$ \\
\hline
\end{tabular}
\end{center}
\end{table}

\subsection{Leptonic Decays}

\begin{figure}[htb]
\begin{center}
\includegraphics*[height=5.5cm]{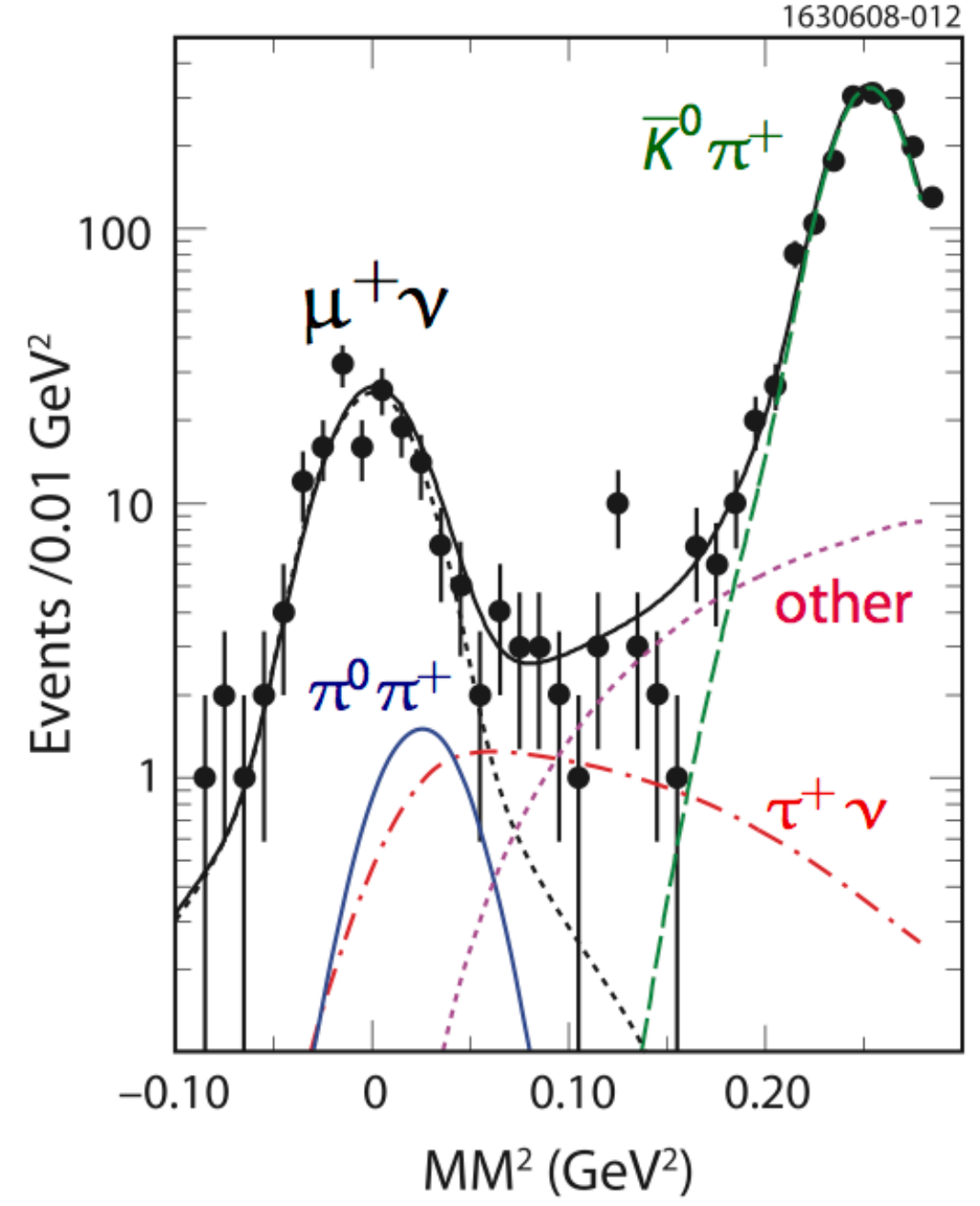}
\includegraphics*[height=5.5cm]{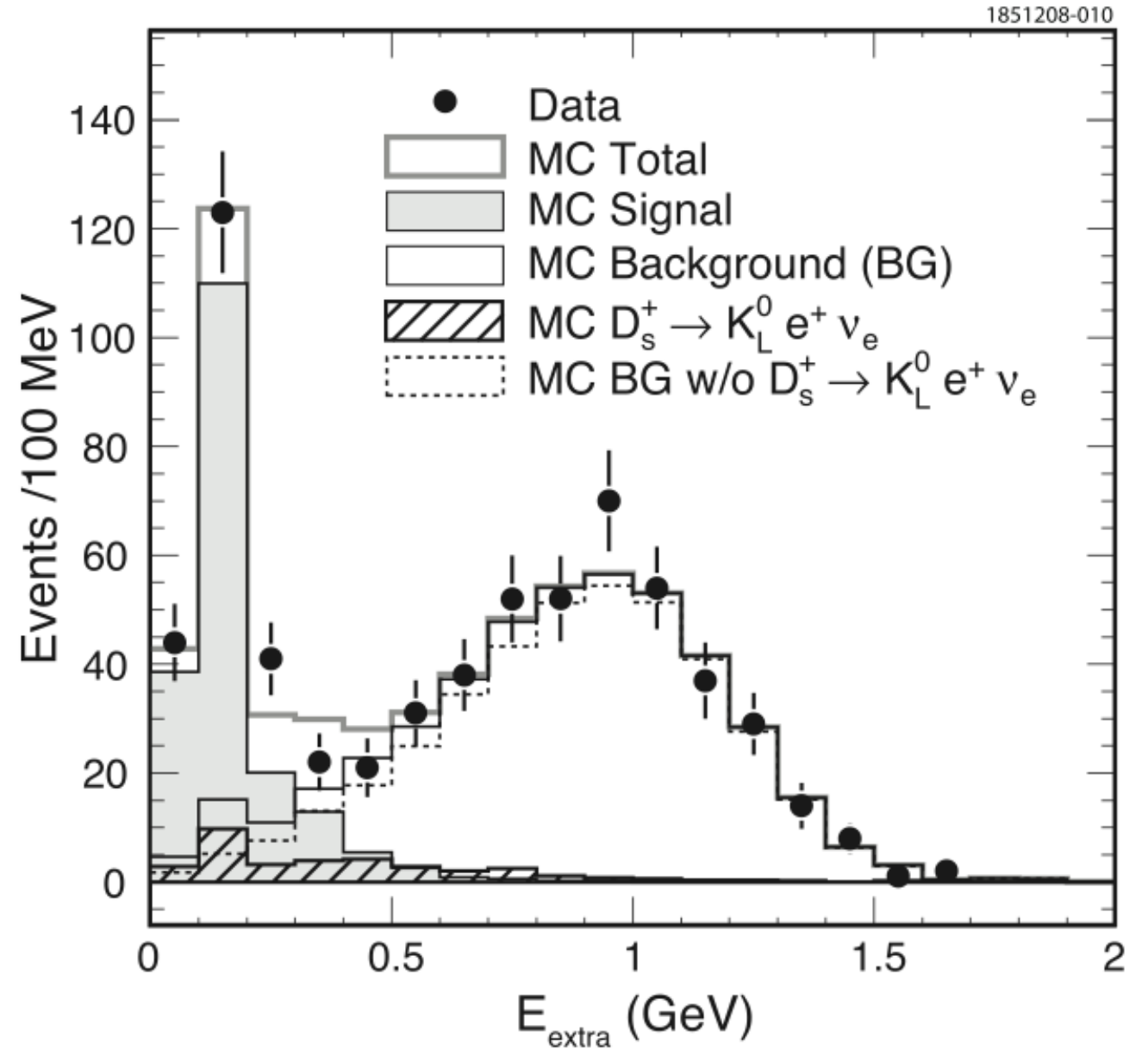}
\includegraphics*[height=5.5cm]{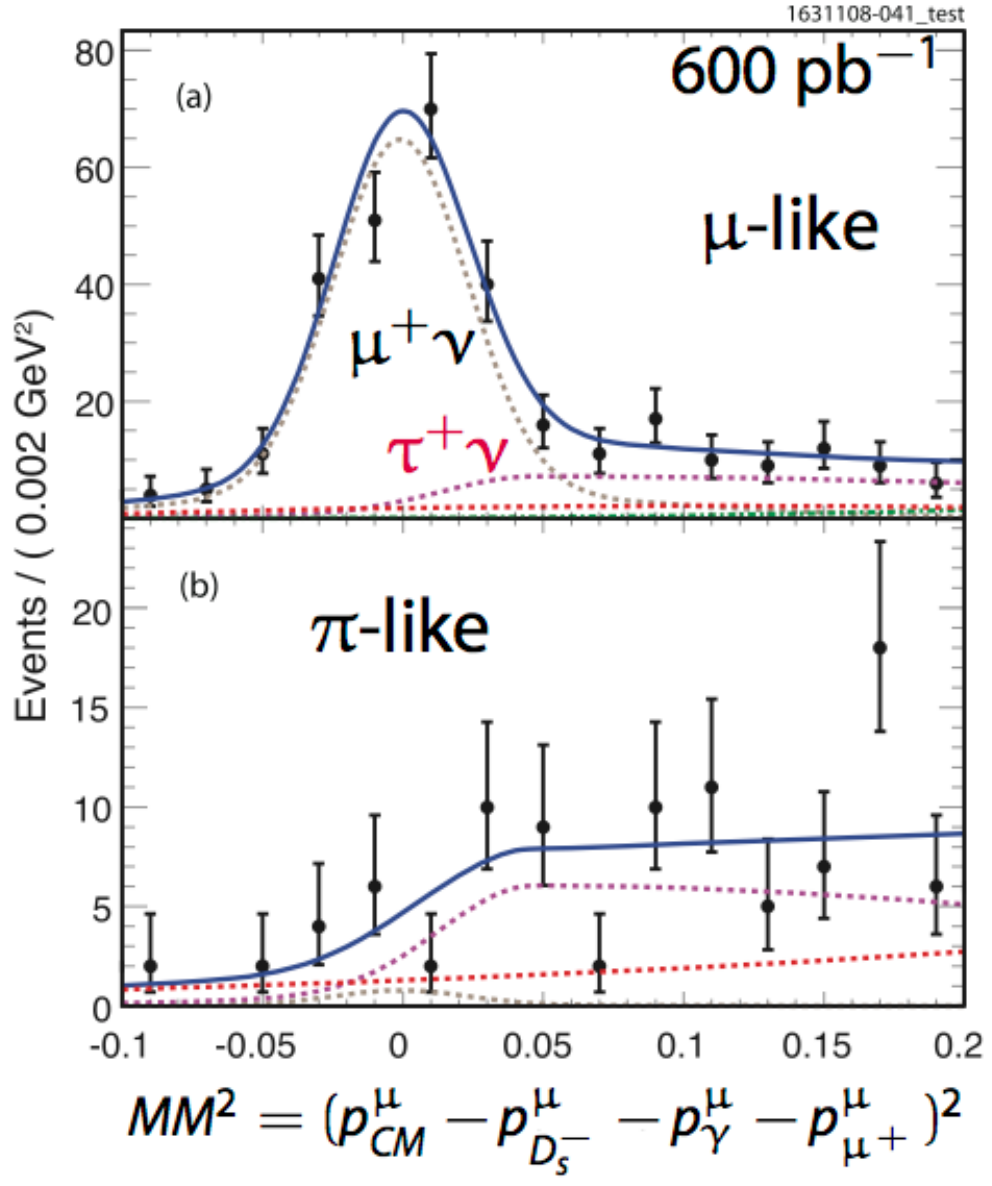}
\caption{Missing-mass-squared distributions for $D^+\to\mu^+\nu$ (left) and
$D_s^+\to\mu^+\nu/\tau^+(\pi^+\bar\nu)\nu$ (right); and unassociated calorimeter
energy for $D_s^+\to\tau^+(e^+\nu\bar\nu)\nu$ (center).
}
\label{fig:lep}
\end{center}
\end{figure}

Results for $D^+\to\mu^+\nu$~\cite{dmunu} and
$D_s^+\to\{\mu^+,\tau^+\}\nu$~\cite{dstaunu,dsmunu} are based on
the full CLEO-c datasets taken at the $\psi(3770)$ and $\sqrt{s}=4170$ MeV,
respectively.
The decay $D^+\to\mu^+\nu$ is both Cabibbo-suppressed and helicity-suppressed.
To search for this rare decay, we combine a tag $D^-$ candidate with a
$\mu^+$ candidate and compute the missing (recoil) mass in the event,
after discarding events with extra tracks and energy deposits in the
electromagnetic calorimeter.
The resultant missing-mass-squared distribution is shown in Fig.~\ref{fig:lep},
and the fitted yield is $149.7\pm 12.0$ events.  No evidence for
$D^+\to\tau^+\nu$ is observed.  The $D^+\to\mu^+\nu$ yield corresponds to
${\cal B}(D^+\to\mu^+\nu)=(3.82\pm 0.32\pm 0.09)\times 10^{-4}$ and
$f_D = (205.8\pm 8.5\pm 2.5)$ MeV.  This measurement of $f_D$ agrees well with
the LQCD calculation~\cite{Follana:2007uv} of $f_D= (207\pm 4)$ MeV.

Unlike $D^+\to\mu^+\nu$, the decays $D_s^+\to\mu^+\nu$ and $D_s^+\to\tau^+\nu$
are Cabibbo-favored and, in the case of $D_s^+\to\tau^+\nu$, not
helicity-suppressed.  To obtain a measurement of $f_{D_s}$, we combine two
analyses.
The first analysis is sensitive to $D_s^+\to\tau^+\nu$, where
$\tau^+\to e^+\nu\bar\nu$.  We select events with a tag $D_s^-$ candidate,
a $e^+$ candidate, and no additional tracks.  Apart from energy deposits
associated with these particles, signal events contain low calorimeter activity.
Fig.~\ref{fig:lep} shows the energy of unassociated calorimeter deposits,
where the displacement of the signal peak from zero arises from the transition
$\gamma$ from the $D_s^*$ decay.  Based on the signal region below 400 MeV, we
obtain
a branching fraction of ${\cal B}(D_s^+\to\tau^+\nu)=(5.30\pm 0.47\pm 0.22)\%$.

The second analysis is a simultaneous treatment of $D_s^+\to\mu^+\nu$ and
$D_s^+\to\tau^+\nu$, where $\tau^+\to\pi^+\bar\nu$.  Here, a tag $D_s^-$
candidate is combined with a track and a photon candidate, and we compute the
missing mass in the event.  Extra tracks and calorimeter energy are vetoed, and
events are classified according to the calorimeter energy matched to the signal
track as either $\mu$-like ($E<300$ MeV) or $\pi$-like ($E>300$ MeV).
Missing-mass-squared distributions for both types of events are shown in
Fig.~\ref{fig:lep}, and we obtain branching fractions of
${\cal B}(D_s^+\to\tau^+\nu)=(6.42\pm 0.81\pm 0.18)\%$ and
${\cal B}(D_s^+\to\mu^+\nu)=(5.65\pm 0.45\pm 0.17)\times 10^{-3}$.

The average $D_s^+$ decay constant from these three branching fraction
measurements is $f_{D_s}=(259.5\pm 6.6\pm 3.1)$ MeV, which also agrees with the
LQCD calculation~\cite{Follana:2007uv} of $f_{D_s}=(241\pm 3)$ MeV.
We also obtain the ratio
$f_{D_s}/f_D = 1.26\pm 0.06\pm 0.02$, which LQCD predicts to be $1.164\pm 0.011$.

\section{Summary}
Data taken at charm threshold provides a unique opportunity to investigate
non-perturbative QCD.
CLEO-c has performed extensive studies of semileptonic and leptonic decays of
$D^0$, $D^+$, and $D_s^+$ mesons.  The measured form factors and decay constants
agree well with new LQCD predictions, which have uncertainties of similar size
to experiment.


\end{document}